\documentclass[epj]{svjour}
\usepackage{amsmath}
\usepackage{graphics}
\begin{document}
\title{Phase separation of the Potts model in the square lattice}
\author{Miguel Ib\'a\~nez de Berganza \inst{1} \and Ezequiel E. Ferrero\inst{2}
\and Sergio A. Cannas \inst{2} \and Vittorio Loreto \inst{1} \and Alberto Petri \inst{3}
}
\institute{Dipartimento di Fisica, Universit\`a di Roma ``La Sapienza''. Piazzale A. Moro 2, 00185 Roma, Italy. \and Facultad de Matem\'atica, Astronom\'ia y F\'isica, Universidad Nacional de C\'ordoba, Ciudad Universitaria, 5000 C\'ordoba, Argentina. \and CNR, Istituto dei Sistemi Complessi, sede di Roma 2-Tor Vegata. Via del Fosso del Cavaliere 100, 00133 Roma, Italy.}

\date{Received: date / Revised version: date}

\abstract{When the two dimensional $q$-color Potts model in the square lattice is quenched at zero temperature with Glauber dynamics, the energy decreases in time following an Allen-Cahn power law, and the system converges to a phase with energy higher than the ground state energy after an arbitrary large time when $q>4$. At low but finite temperature, it cesses to obey the power-law regime and orders after a very long time, which increases with $q$, and before which it performs a domain growth process which tends to be slower as $q$ increases. We briefly present and comment numerical results on the ordering at nonzero temperature.
\PACS{
      {PACS-key}{discribing text of that key}   \and
      {PACS-key}{discribing text of that key}
     } 
} 

\maketitle

The theory of ordering dynamics \cite{Lifshitz,Allen,Bray,Gunton} concerns the dynamic evolution of a system which coarsens from one disordered equilibrium phase to an ordered state, when it is quenched at a temperature well deep inside the ordered phase. Systems with scalar order parameter, as the Ising model, can be studied with the Allen-Cahn theory \cite{Bray}, which consists in an analysis of a Time-Dependent Ginzburg Landau equation of motion for the order parameter. The theory postulates the validity of the so called \textit{scaling hypothesis}, which states that the system is statistically invariant in time when space is rescaled by a time-dependent characteristic lenght, $\ell$, proportional to the mean size of the growing domains. Through the scaling hypothesis, and for Ising-like systems, the Allen-Cahn theory predicts a power-law growth in time, the \textit{Allen-Cahn law}, for the characteristic domain length, $\ell \sim t^{\alpha}$, being $\alpha=1/2$ for non-conserved order parameter systems \cite{Bray}, law which has been verified numerically for both continuous and lattice Ising models \cite{RefsIsing}. \\
For systems with a $q>2$-times degenerated ground state, as the $q$-state Potts model, the situation is less understood \cite{Lifshitz,Bray}: the Allen-Cahn theory predicts also in this case an Allen-Cahn power law, provided that the scaling hypothesis is satisfied. After some initial controversy, the Allen-Cahn law with exponent $\alpha=1/2$ was finally found solving numerically the dynamics of the Potts model in the lattice \cite{Grest88,Jeppesen,Potts-numeric}, and the scaling hypothesis has also been verified in a given temporal range \cite{Lau,Jeppesen,Sire}, when domains are still not comparable with the system size.\\
On the other hand, recent studies has shown an interesting behavior when the $q>4$ Potts model in the square lattice is quenched at zero temperature \footnote{See \cite{IbanezPM,future} for a detailed review on these findings, in which the relation of the problem with the order of the thermodynamic and infinite-time limits is dicussed and explained.} \cite{IbanezPM,PetriEL,Petri}: in this case, the system converges after an infinite time to a disordered, stationary phase with nonzero energy density, $e^*(q)>0$ for $q>4$, different from the equilibrium state at zero temperature, which was called ``glassy phase'' in \cite{PetriEL}. Moreover, in these works it is shown how the ordering of the system can be described, after a short initial period, by a power-law for the energy, $e(t,q)=e^*(q)+a(q)\ t^{-1/2}$. Being the exess energy at low temperature equal to the perimeter of the interface, it is $e\sim \ell^{-1}$ and, hence, the above result implies that the domains do not grow indefinitely, but the characteristic scale $\ell$ converges to a limit value, $\ell^*(q)$. The Allen-Cahn law is, in this way, not strictly respected, even if, during the coarsening, it is $e(t,q)\sim t^{-1/2}$.\\
For what concerns the ordering in the presence of thermal fluctuations, it has been recently found for $q=7$ \cite{IbanezPM} that at low but positive temperature, $T=0.1$, the model orders in such a way that the non-homogeneous Allen-Cahn law, $e(t,q)=e^*(q)+a(q)\ t^{-1/2}$, valid at zero temperature, is respected with same functions $e^*(q)$ and $a(q)$, but only up to a time, $\tau(T,q)$, in which the system ceases to obey the power law, and after which the mean domain size converges very fast to the size of the system, and the energy to zero (see details in \cite{IbanezPM}). Figure 1 shows this phenomenology: we report the energy per site versus time of the kinetic 2d-Potts model in the square lattice with periodic bonduary conditions, when the system evolves with single-spin flip Glauber dynamics after a quench to $T=0.1$, being $T_c(q)=\ln^{-1}(1+q^{1/2})$ the critical temperature of the model \cite{Wu,Baxter}, which takes the values $0.482...$ for $q=48$ and $0.994...$ for $q=3$. These critical temperatures correspond to a continous phase transition for $q\le 4$, and to a first-order transition for $q>4$, and this $q$-range coincides with the one in which $e^*(q)>0$
. We show in the figure the $q=3,9,20,48$ cases, for a system of linear size $L=200$ (see \cite{IbanezPM} for a finite-size effect analysis). We see that the $q=3$ case exhibits a power law, and $e^*(q=3)=0$. For $q>4$, it is $e^*>0$, and the transient time after which the system exhibits the power-law relaxation, say $\tau_i(q)$, increases with $q$. It was shown to be \cite{IbanezPM} of the order of 10 Monte-Carlo steps (MCS) for $q=7$, but it is roughly $10^3$ MCS for $q=48$. \\
We now breafly point out some other characteristics of the data in Figure 1, characteristics which are extensively exposed and interpreted in terms of the Allen-Cahn theory and the scaling hypothesis in \cite{future}, where we will also discuss the influence of pinned local structures in the problem. \textit{Supposing the non-homogeneous Allen-Cahn law above defined to be satisfied in the interval} $\tau_i(q)<t<\tau(q)$ (and we arbitrarely define the interval as the one in which a power-law can be identified), we have four parameters characterizing the ordering of the energy: $\tau(q,T)$, $\tau_i(q)$, $a(q)$ and $e^*(q)$. From the analisys of Figure 1 we can notice some aspects of these parameters: \textbf{1)} first of all, the fact that $e^*(q)$ increases with $q$, as was already exposed in \cite{Petri}, where it was also found numerically the behaviour $e^*(q)=b\ (q-4)^{1/2}$ for $q$ from 2 to 20 (see also \cite{IbanezPM} for a discussion on this point). We find indeed the same pre-factor $b=0.06$ found for the zero-temperature dynamics in \cite{Petri}. 
\textbf{2)} $\tau(T,q)$ increases with $q$; the larger the number of colours, the less probable to find the fluctuations that will lead to the formation of domains of the size of the system \cite{IbanezPM}. \textbf{3)} $\tau_i(q)$, the limit for the transient regime, increases with $q$, as already mentioned above. \\
To summarize, we have described some of the recent results on the ordering of the 2d-Potts model in the square lattice, and presented new results on the ordering at positive temperature. At zero temperature the system converges to a phase with nonzero energy density for $q>4$, while at low temperature it escapes from this phase in a time, $\tau$, which increases with $q$. The ordering before $\tau$ is such that the energy decreases with the Allen-Cahn power law with exponent 1/2, but the pre-factor is very low for large $q$. A detailed analysis of these data together with an interpretation of them in terms of the Allen-Cahn theory and the scaling hypothesis is being performed \cite{future}. Study of the temperature dependence of the problem and of the influence of pinned local structures \cite{Spirin,Godreche,Derrida96} is also in progress.

\begin{figure}
\begin{center}
\resizebox{0.5\textwidth}{!}{
 \includegraphics{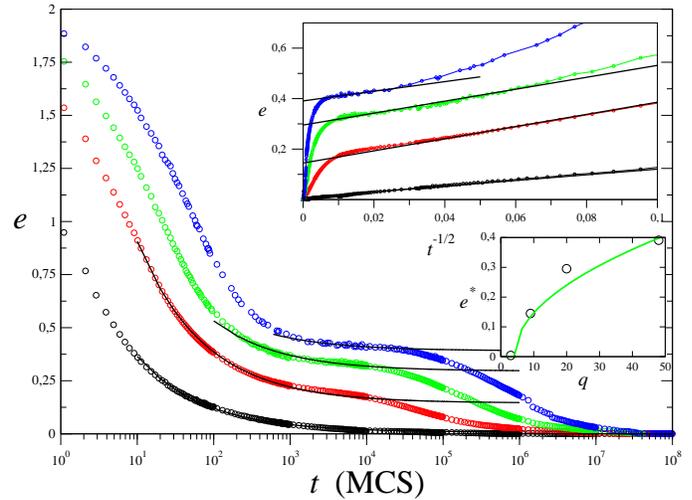}
}
\caption{Energy per site ($e$) vs. time ($t$) for $q=3,9,20,48$ (in increasing energy order). On upper inset, $e$ vs. $t^{-1/2}$ and the fit with the non-homogeneous Allen-Cahn law $e(t,q)=e^*(q)+a(q)\ t^{-1/2}$ (see text). On lower inset, the fitted $e^*(q)$ (circles), and the function $b (q-4)^{-1/2}$, $b=0.06$ found in \cite{PetriEL} for zero-temperature data (line).}
\label{fig:1}       
\end{center}
\end{figure}


\end{document}